\documentclass{phb-proc4-auth}
\usepackage{graphicx}
\usepackage{amssymb}

\begin{document}
\begin{frontmatter}

\journal{SCES '04}

\title{Variations of Pairing Potential and Charge 
Distribution
in Presence of a Non-magnetic Impurity}

\author[Dr,Lu]{Grzegorz Litak \thanksref{E-mail}}

\address[Dr]{Max Planck Institute for Physics of Complex Systems,
 N\"{o}thnitzer Str. 38, D--01187 Dresden, Germany
}
\address[Lu]{
Department of Mechanics, Technical University
of Lublin,
Nadbystrzycka 36, PL-20-618 Lublin, Poland}

\thanks[E-mail]{Tel.: +48- 81- 5381573; Fax: +48- 81- 5241004; E-mail:
g.litak@pollub.pl}

\corauth[]{}

\begin{abstract}
Using an attractive Hubbard model we examine spatial
variations of superconducting order parameter
and local charge 
on a two dimensional lattice.
 For various band filling we show
the effect of destruction of the order parameter around a non-magnetic 
impurity. In case of  a half-filled system such destruction is accompanied 
by appearance of characteristic charge variations around the impurity
with an isotropic 
distribution. 
\end{abstract}

\begin{keyword}
superconductivity, non-magnetic impurities, charge density wave
\end{keyword}

\end{frontmatter}

According to  Anderson theorem the effect of non-magnetic disorder on
s-wave superconductors can be neglected unless
spatial
fluctuations of order parameter are present \cite{And59,Gyo96}. 
However this 
theorem invented for
conventional superconductors
cannot be applied  to other superconductors with  a short
coherence
length 
\cite{Lit98,Mor01} and for those of anisotropic pairing 
\cite{Gor83}.
Another interesting situation, where
the influence of disorder can be important,  
has been found in a case with  an 
interplay 
between different long range orders like  superconductivity 
and charge density wave (CDW). In a negative $U$ Hubbard model this 
happens 
for 
half-filling   \cite{Mil93}. Here finite disorder favours 
superconducting phase  against CDW
\cite{Lit98,Hus97,Hus98,Paw01}. These results 
show 
that the effect of disorder on superconductivity can be 
sometimes beneficial  
leading to disorder induced superconductivity \cite{Paw01}. It must be 
also noticed that the charge 
ordering mechanism and its interplay with  superconductivity are of 
a great interest itself because of existence of  charge strips in HTc 
superconductors \cite{Tim99}. 
In the present note we will examine the
effect of a single impurity on a superconducting order parameter 
and charge fluctuations around non-magnetic impurity in a two dimensional 
lattice.

\begin{figure}[htb]
\begin{center}
~
\vspace{-0.6cm}

\includegraphics[angle=-90,width=16.5pc]{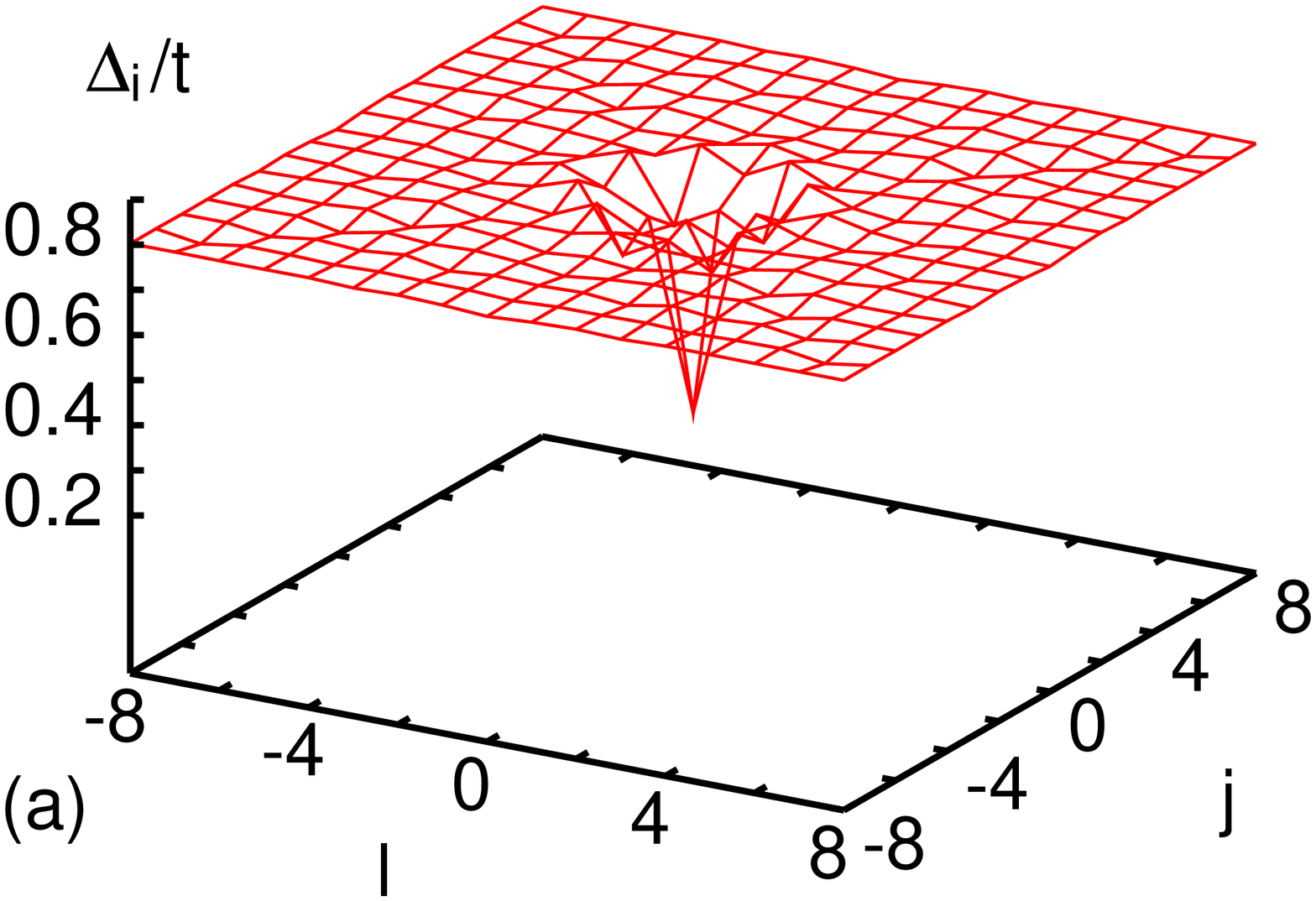}
\vspace{-0.6cm}

\includegraphics[angle=-90,width=16.5pc]{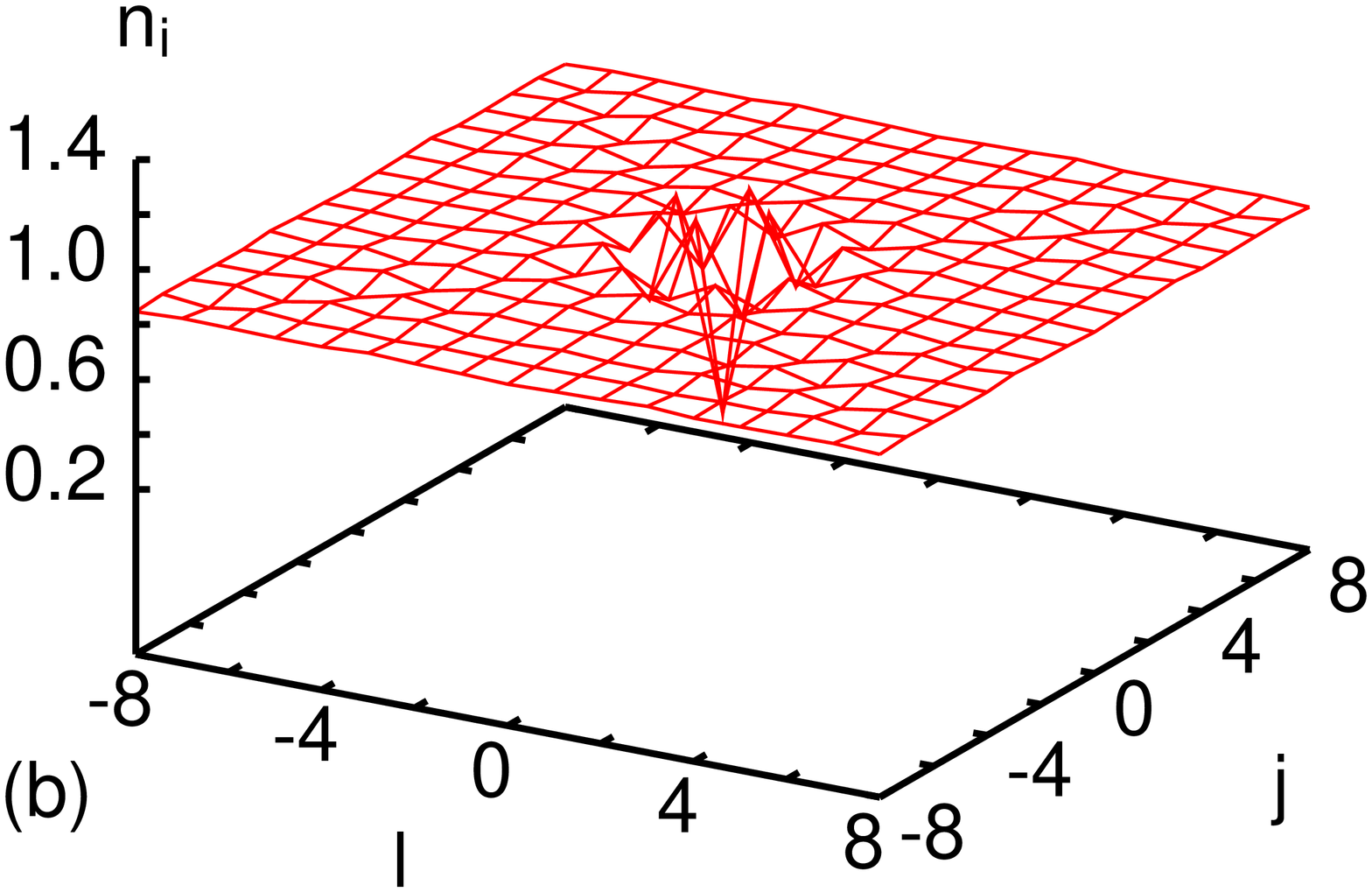}

\end{center}  
\caption{ \label{fig:1}
Pairing potential $\Delta_i$ (a) and charge $n_i$ ($i=<l,j>$) on a two  dimensional
lattice around the central impurity ($U=-3t$, $n=0.85$,  $\epsilon_0=t$
at  $(l,j)=(0,0)$ ).}
 \end{figure}

We start from negative $U$ Hubbard hamiltonian \cite{Lit98,Mor01,Gho02}:
\begin{eqnarray}
H &=&-t\sum_{\langle i,j \rangle \sigma} \left(c_{i \sigma}^\dagger c_{j 
\sigma}+h.c.\right)+
  \sum_{i \sigma} \left(\epsilon_i-\mu \right) c_{i \sigma}^\dagger c_{i 
\sigma} \nonumber \\ &+&
  U \sum_{i} n_{i \uparrow} n_{i \downarrow},
\label{hamiltonian}
\end{eqnarray}
where
$\epsilon_i$  denotes  an impurity 
potential located at the central site $i=0$ ($\epsilon_i =\epsilon_0 
\delta_{0,i}$) and $U$  
on--site attraction
($U < 0$) which is the same at each lattice site $i$.

Our actual calculations consist of solving, self-consistently,
the following Bogoliubov-de Gennes equation  \cite{Gen66}:
\begin{eqnarray}
 \sum_{j} \left(\begin{array}{c}
 (E^\nu - \varepsilon_i + \mu)\delta_{ij} + t_{ij} ~ ~ ~ ~
 \Delta_i \delta_{ij} \\
 \Delta_i^* \delta_{ij} ~ ~ ~ ~
 (E^\nu +  \varepsilon_i - \mu) \delta_{ij}  - t_{ij}
\end{array}\right)
\left(\begin{array}{ll}
 u^\nu_{j}\\
v^\nu_{j}\end{array}\right)=0 \,, \nonumber \\
\label{bogoliubov}
\end{eqnarray}
where  $ \varepsilon_i= \epsilon_i+ U n_i/2$ denotes the renormalized site 
energy.
The pairing potential $\Delta_i$ and the local charge $n_i$ are 
to be found
self-consistently:
\begin{eqnarray}
\Delta_i &=&
-U \sum_{\nu} u^\nu_{i}v^{\nu*}_{i}
(1 - 2f(E^\nu))\,,   \nonumber \\
n_i &=& 2 \sum_{\nu} (|u^\nu_{i}|^2 f(E^\nu)+|v^\nu_{i}|^2         
(1 - f(E^\nu)))\,,
\label{selfcons}
\end{eqnarray}
where $\nu$ enumerates the solutions of Eq.~\ref{bogoliubov}
for a given band filling $n$.

\begin{figure}[htb]
\begin{center}
~
\vspace{-0.6cm}

\includegraphics[angle=-90,width=16.5pc]{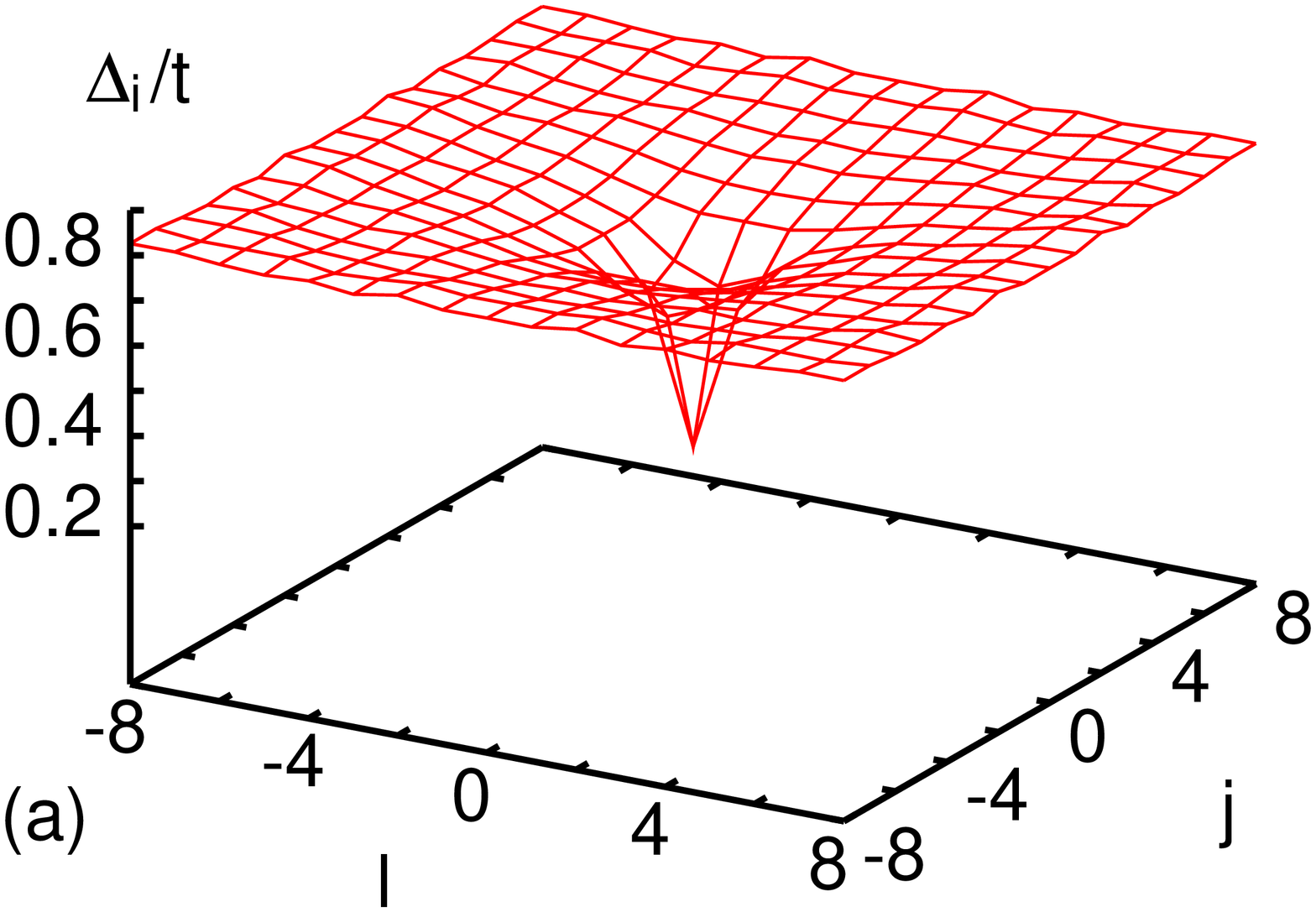}
\vspace{-0.6cm}

\includegraphics[angle=-90,width=16.5pc]{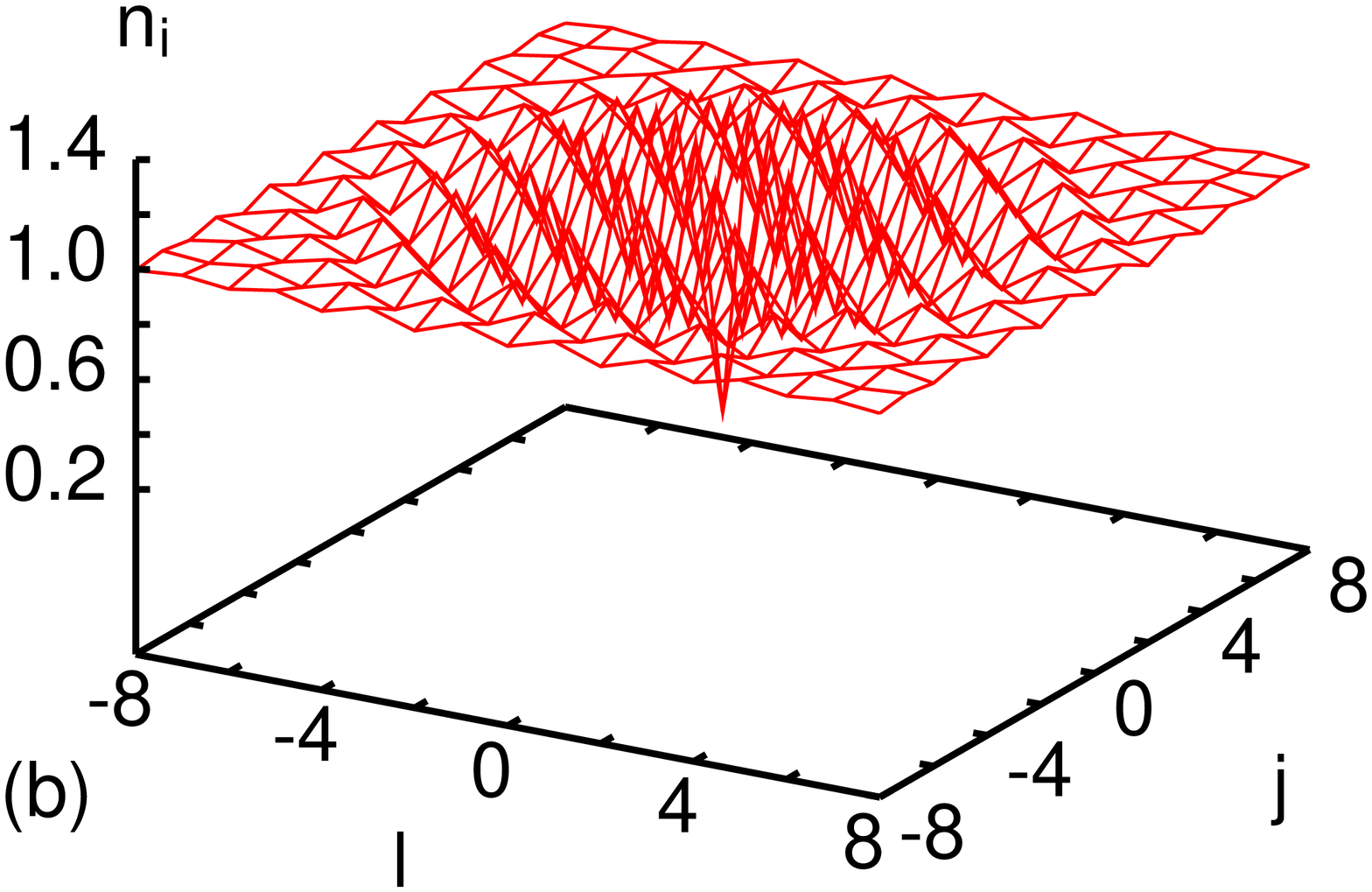}

\end{center}
\caption{ \label{fig:2}
Pairing potential $\Delta_i$ (a) and charge $n_i$ ($i =<l,j>$) on a two  
dimensional 
lattice around the central impurity ($U=-3t$, $n=1.0$,  $\epsilon_0=t$
at  $(l,j)=(0,0)$ ).}  
 \end{figure}

To examine the effect of a single impurity on the surrounding lattice 
sites
we have solved the above equations (\ref{bogoliubov},\ref{selfcons}) in
the real space using the recursion Lanczos algorithms for a
superconductor \cite{Lit95}.
In Fig. \ref{fig:1} we show the local distributions of  the pairing 
potential 
$\Delta_i$ 
(Fig. 
 \ref{fig:1}a) and 
the charge $n_i$ (Fig.  \ref{fig:1}b) around impurity located in the 
center of a 2d 
lattice for a 
band 
filling $n$ close but slightly  smaller than a half-filed situation 
($n=0.85$). 
Due to the impurity ($\epsilon_0=t$)
the pairing potential $\Delta_i$ is going down rapidly at the central site 
(Fig.  
\ref{fig:1}a). 
This change is 
coupled to variation of the local charge $n_i$ (Fig.  \ref{fig:1}b).
Note that both distributions of $\Delta_i$ and $n_i$  have similar form
e.g. they show similar anisotropy.  
The situation is
quite different for a half filled system $n=1$  (Fig.  \ref{fig:2}).  
Note that in this case the paring potential goes down around impurity more 
smoothly and isotropically (Fig.  \ref{fig:2}a) in comparison to the 
previous case (Fig.  
\ref{fig:1}a). Interestingly,  
destruction of pairing is associated with 
strong oscillation of charge $n_i$ (Fig.  \ref{fig:2}b). It is clear that 
around the 
central 
impurity the electron charge form a localized wave, with a characteristic 
size of 
8-9 lattice spaces.  In case of finite concentration of impurities we can 
expect superconductor with  islands of a normal phase \cite{Gho02}.  Our 
calculations 
show 
the appearance of local CDW. That in turn indicates that impurities can 
contribute to 
a  phase separation phenomenon \cite{Paw01}.  \\ \\ 
{\bf Acknowledgements} \\
This work has been partially supported by the KBN grant No. 2P03B06225.

%
%
%
%


\end{document}